\documentclass[12pt]{article}
\begin{document}

\title{{\bf Predictions and Tests
\\ of Multiverse Theories}
    \thanks{Alberta-Thy-13-04, hep-th/0610101, to appear in
B.~J.~Carr, ed., {\em Universe or Multiverse?} (Cambridge University
Press, Cambridge, 2007), pp. 401-419.}}

\author{
Don N. Page
\thanks{Internet address:
don@phys.ualberta.ca}
\\
Institute for Theoretical Physics\\
Department of Physics, University of Alberta\\
Room 238 CEB, 11322 -- 89 Avenue\\
Edmonton, Alberta, Canada T6G 2G7
}

\date{(2004 June 25; edited 2004-2006 by Bernard Carr)}

\maketitle
\large

\begin{abstract}
\baselineskip 18 pt

Evidence for fine-tuning of physical parameters suitable for life can
perhaps be explained by almost any combination of providence,
coincidence or multiverse. A multiverse usually includes parts
unobservable to us, but if the theory for it includes suitable
measures for observations, what is observable can be explained in
terms of the theory even if it contains such unobservable elements.
Thus good multiverse theories can be tested against observations.  For
these tests and Bayesian comparisons of different theories that
predict more than one observation, it is useful to define the concept
of ``typicality'' as the likelihood given by a theory that a random
result of an observation would be at least as extreme as the result of
one's actual observation. Some multiverse theories can be regarded as
pertaining to a single universe (e.g.\ a single quantum state obeying
certain equations), raising the question of why those equations
apply.  Other multiverse theories can be regarded as pertaining to no
single universe at all. These no longer raise the question of what the
equations are for a single universe but rather the question of why the
measure for the set of different universes is such as to make our
observations not too atypical.

\end{abstract}
\normalsize
\baselineskip 15.8 pt
\newpage

\section{Multiverse explanations for fine-tuning}

Many of the physical parameters of the observed part of our universe,
whether constants of nature or cosmological  boundary conditions, seem
fine-tuned for life and us \cite{Carter,Carr-Rees,Barrow-Tipler,Rees}.
There are three common explanations for this.  One is that there is a
Fine-Tuner who providentially selected the physical parameters so that
we can be here. Another is that it is just a coincidence that the
parameters turned out to have the right values for us to be here. A
third is that our observed  universe is only a small part of a much
vaster universe or multiverse or megaverse or holocosm (my own
neologism for the whole), and that the physical parameters are not the
same  everywhere but take values permitting us in our part.

These three explanations are not necessarily mutually exclusive.   For
example, combining a Fine-Tuner with coincidence but without a
multiverse, perhaps the universe was providentially created by a God
who had a preference for a particularly elegant single universe which
only coincidentally gave values for the physical parameters that
allowed us to exist.  Or, for a Fine-Tuner with a multiverse but
without coincidences, perhaps God providentially created a multiverse
for the purpose of definitely creating us somewhere within it.  Or,
for coincidence and a multiverse without a Fine-tuner, if the universe
weren't providentially created, it might be a multiverse that has some
parts suitable for us just coincidentally.  Or, it might even be that
all three explanations are mutually true, say if God providentially
created a multiverse for reasons other than having us within it, and
yet it was a coincidence that this multiverse did contain us.

On the other hand, it seems conceivable (in the sense that I do not
see any obvious logical contradiction) that the universe is determined
by some sort of blind necessity that requires both our own existence
and a single world with a single set of physical parameters. In this
case, the universe is not providential (in the sense of being foreseen
by any God) but nor is our existence coincidental.

Thus, logically, I don't see that we  can prove that any combination
of the three explanations is either correct or incorrect.  However, it
does seem a bit implausible that none of these explanations is at
least partially correct, and it also seems rather implausible that the
large number of fine-tunings that have been noticed are mere
coincidences.

I should perhaps at this point put my metaphysical cards on the table
and say that -- as an evangelical Christian -- I do believe the
universe was providentially created by God, and that -- as quantum
cosmologist with a sympathy toward the Everett `many-worlds' version
of quantum theory -- I also strongly suspect that the universe is a
multiverse, with different parts having different values of the
physical parameters.  It seems plausible to me that -- in a quantum
theory with no arbitrary collapses of the wavefunction -- God might
prefer an elegant physical theory (perhaps string/M theory with no
adjustable dimensionless parameters) that would lead to a multiverse
that nevertheless has been created providentially by God  with the
purpose of having life and us somewhere within it.

Although personally I have less confidence in string/M theory than in
either providence or the multiverse, nevertheless string/M theory is
very attractive. It does seem to be the best current candidate for a
dynamical theory of the universe (i.e.\ for its evolution, if not its
state) and it does strongly appear to suggest a multiverse.  Since
string/M theory has no adjustable dimensionless constants, if it
predicted just a single set of parameters, it would seem  very
surprising if these parameters came out right for our existence. Thus
if string/M theory -- or some alternative with no adjustable
dimensionless constants -- were correct, it would seem much more
plausible that  it would lead to a multiverse, with different parts of
the universe having different physical parameters.

Indeed, string theorists \cite{DNP,BP,KKLT,Suss,Doug,BDG,FST} have
argued that string/M theory leads to an immense multiverse or
landscape of different values of physical parameters and `constants of
nature'. It is not yet known whether the range of values can include
the physical parameters that allow life, such as those within our part
of the universe, but that does seem at least plausible with the
enormous range suggested in the string landscape or stringscape.

One objection that is often raised against the multiverse is that it
is unobservable.  Of course, this depends on how the multiverse is
defined.  One definition would be the existence of different parts,
where  some physical parameters are different, but this just shifts
the arbitrariness to the choice of this set of physical parameters. 
Obviously if some quantity which varies with position (like energy
density) were included in the set of physical parameters, then even
what we can see could be considered a multiverse. But if we just
include  the so-called `constants of nature', like the fine structure
constant and various other coupling constants and the mass ratios of
the various elementary particles, then what we can observe directly
seems to consist of a single universe. Indeed it would be rather
natural -- if {\it ad hoc} -- to define a multiverse with respect to
the physical parameters that have no observable variation within the
part we can directly see.   In this case,  the multiverse becomes
unobservable, and it becomes an open question whether parts of the
universe we cannot see have different values of these constants. Many
would argue that it is a purely metaphysical concept that has no place
in science. 

However, in science we need not restrict our entities to be observable
-- we just want the simplest theory, whether using observable or
unobservable entities, to explain and predict what is observable.  One
cannot test scientifically a theory that makes predictions about what
is unobservable, but one can test a theory that makes use of
unobservable entities to explain and predict the observable ones.
Therefore, if we find a multiverse theory that is simpler and more
explanatory and predictive of what is observed than the best
single-universe theory, then the multiverse theory should be
preferred.  The success of such a multiverse theory itself would then
give credence to the existence of the unobservable multiverse.

Another objection that is often raised against multiverse theories is
that na\"{\i}vely they can `explain' anything and predict nothing, so
that they cannot be tested and  considered scientific.  The idea is
that if a multiverse gives all possible physical parameters or other
conditions somewhere within the multiverse, then the parameters and
conditions we observe will exist somewhere. Hence what we observe is
`explained' at least somewhere.  On the other hand, if every
possibility exists, then we cannot predict any non-trivial restriction
on what might be observed. If a theory makes no non-trivial
predictions, then it cannot be tested against observations, and it can
hardly be considered  scientific.

\section{Testable multiverse explanations}

Sufficiently sophisticated multiverse theories can provide predictions
as well as explanations, and hence can be tested against observations
scientifically.  Unlike single-universe theories, in each of which one
can in principle predict uniquely the physical parameters, in
multiverse theories one usually can make only statistical predictions
for ranges of parameters, but this can still be much better than
making no prediction at all. However, to make such statistical
predictions, the multiverse theory needs to include a {\it measure}
for the different observations that can be made.  If it allows all
possible observations without putting any measure on them, then one
can make no predictions.

Since we have strong evidence that we live in a quantum universe, it
would be natural to seek a quantum multiverse theory.   If this just
includes some quantum states, unitary evolution, path integrals,
operators, some operator algebra and the like, one has the bare
quantum theory eloquently described by Sidney Coleman \cite{Cole},
which by itself does not give any measures or probabilities.  The
Copenhagen version of quantum theory does give these, but at the
apparent cost of the collapses of the wave-function at times
undetermined by the theory and to states that are random. 

Here I shall take essentially an Everett `many-worlds' view that in
actuality there is  no collapse of the wavefunction.  However, to get
testability of the quantum theory, I shall assume that there is one
aspect of the Copenhagen version that should be added to the bare
quantum theory: measures for observations that are expectation values
of certain corresponding `awareness operators'.

In Copenhagen theory, these operators are projection operators, and
the measures are the probabilities for the results of the collapse of
the wave-function.  Here I shall not necessarily require the operators
to be projection operators, though -- to give the positivity
properties of measures -- I shall assume they are at least positive
operators.  Also, I shall not assume that  anything really random
occurs, such as wave-function collapse, but that there are simply
measures for all the different observations that might occur.  In
testing the theory against one's observation, one can regard that
observation as being selected at random (with the theory-given
measure) from the set of all possible observations, but ontologically
one can assume that all possible observations with non-zero measure
really do occur, so that there is never a real physical random choice
between them.

For the quantum theory to be fundamental, one would need to specify
which observations have measures and what the corresponding operators
are whose expectation values give those measures.  In my opinion, the
most fundamental aspect of a true observation is a conscious
perception or awareness of the observation. Therefore, I have
developed the framework of ``Sensible Quantum Mechanics'' (SQM)
\cite{SQM} or ``Mindless Sensationalism'' \cite{MS} for giving the
measures for sets of conscious perceptions as expectation values of
corresponding positive operators that I call ``awareness operators''.
This is only a framework (analogous to the bare quantum theory without
the detailed form of the unitary evolution or operator algebra),
rather than a detailed theory, since I have no detailed proposal for
the sets of possible conscious perceptions or for the corresponding
positive operators.  Presumably, for human conscious perceptions,
these operators  are related to states of  human brains, so
understanding them better would involve brain physics. However, I do
not see how they could be deduced purely from an external examination
of a brain, since we cannot then know what is being consciously
experienced by the brain.

To avoid the complications of brain physics, one might use the
observed correlation between external stimuli and conscious
experiences to replace the unknown awareness operators acting on
brain-states  with surrogate operators acting on the correlated
external stimuli.  Of course, this would not work well for illusory or
hallucinatory conscious perceptions, for which the fundamental
awareness operators would presumably still work if they were known.
However, one might prefer to focus on conscious perceptions that are
correlated with external stimuli and hence better fit what is usually
meant by observations.

If the awareness operator for a conscious perception is correlated
with a single set of external stimuli at a single time, it could be
approximately replaced with a single projection operator onto some
external system.  Alternatively, if it is correlated with a sequence
of measurement processes, then it could be approximately replaced with
a product of projection operators or a sum of such products, a class
operator of the decoherent histories approach to quantum theory
\cite{Grif,GMH,Omnes}.

Therefore, though I would not regard either the projection operators
of Copenhagen quantum theory or the class operators of decoherent
histories quantum theory as truly fundamental in the same way that I
believe awareness operators are, it might be true that in certain
circumstances these are reasonable approximations to the fundamental
awareness operators. Then one can take their expectation values in the
quantum state of the universe as giving the measure for the
corresponding conscious perception.

One example of this replacement would be to calculate the measure for
conscious perceptions of a certain value of the Hubble constant.  In
principle, in SQM this would be the expectation value of a certain
awareness operator that presumably acts on suitable brain-states in
which the observer is consciously aware of that particular Hubble
constant value. But the expectation value of this operator might also
be well approximated by that of some suitable operator acting on the
logarithm of the expansion rate of the part of the universe that is
observed. Because the latter operator does not involve brain physics,
it might be easier to study scientifically and so could be used as a
good surrogate for the actual awareness operator.

However, it would presumably not be a good approximation to use the
latter operator if its expectation value depended significantly on
parts of the universe where there are no conscious observers:  if one
wants to use it to mimic the expectation value for the perceptions of
conscious observations, one must include a selection effect which
restricts to parts of the universe where there are conscious
observers.

To include this selection effect in operators that are external to
brains (or whatever  directly has the conscious perceptions), so that
their expectation values can be good approximations for that of the
fundamental awareness operators, is a difficult task, since we do not
know the physical requirements for conscious observers.  For example,
there is nothing within our current understanding of physics that
would tell us whether or not some powerful computer is conscious,
unless one makes assumptions about what is necessary for
consciousness. Also, I know of nothing within our current
understanding of physics that would enable us to predict that I am
currently conscious of some of my visual sensations but not of my
heartbeat, since presumably information about both is being processed
by my brain and would be  incorporated in a purely physical analysis.

Nevertheless, to get some very crude guess for a selection effect for
conscious observers, one might make the untested hypothesis that
typical observers are like us in requiring suitable complex chemical
reactions and perhaps a liquid compound like water. Then  one could
use the existence of liquid water as a very crude selection effect for
observers and attach it onto other projection or class operators used
to approximate some conscious perception depending on the external
stimuli that are described by the projection or class operators.

Thus one might use projection or class operators to ask the following
two questions: Does liquid water exist in part of the universe?  Is
that part of the universe expanding at a suitable logarithmic rate? If
the answer to these questions is yes with some measure, then one might
expect that there would be a roughly corresponding expectation value
for the awareness operator for conscious perceptions of that value of
the Hubble constant.  This is an extremely crude approximation to what
I postulate would objectively exist as the expectation value of the
true corresponding awareness operator, but since these awareness
operators are as yet largely unknown, the crude approximation may be
useful during our present ignorance.

One problem with calculating the measure for sets of conscious
perceptions as expectation values of corresponding `awareness
operators' is that na\"{\i}vely one might get infinite values.  By
itself this would not necessarily be a problem, since only ratios of
measures are testable as conditional probabilities.  However, when the
measures themselves are infinite, it is usually ambiguous how to take
their ratios.

The problem arises if the awareness operators are sums of positive
operators that are each localized within finite spacetime regions (as
one would expect if the operators correspond to finite conscious
beings).  Assume that one such operator in the sum has support within
one of $N$ spacetime regions of equal volume within the total
spacetime. Then by translational or diffeomorphism invariance, one
would expect the sum of the operators for a particular awareness
operator to include a sum over the corresponding operators in each of
the $N$ regions.  (There would also be a sum over operators that
overlap different regions, but we need not consider those for this
argument.) This is essentially just the assumption that, if a suitable
brain-state for some conscious perception can occur in one of the $N$
spacetime regions, then it can also occur (depending on the quantum
state) in any of the other $N-1$ regions. Also where it occurs in some
coordinate system should not affect the content of the conscious
perception produced by the corresponding brain-state.

If the conditions for observers with the corresponding conscious
perception occur within all $N$ spacetime regions, so that the
expectation value of the operator within each region has a positive
expectation value bounded from below by a positive number $\epsilon$,
then the total awareness operator (a sum of at least the individual
positive operators within each of the $N$ regions) will have an
expectation value at least as large as $N\epsilon$. This is infinite
if the number $N$ of such spacetime regions is infinite. Essentially
the argument is that, if the measure for a conscious perception has a
strictly positive expectation value for each spacetime volume in some
region, then for an infinite volume of spacetime where this is true
the measure will be infinite.  One can regard this as arising from the
infinite number of conscious observers that arise in an infinite
volume of spacetime with conditions suitable for life and conscious
observers.

Since inflation tends to produce a universe that is arbitrarily large
(with an infinitely large expectation value for the spatial volume at
any fixed time after inflation and hence presumably infinitely many
conscious observers), it tends to produce an infinite measure for
almost all non-zero sets of conscious observations.  There has been a
lot of discussion in the literature \cite{LM,Vil,VVW,GV} of how to get
well-defined ratios of these infinite measures (or of related
quantities, since the discussion is not usually in terms of measures
for conscious perceptions) but I think it is fair to say that there is
as yet no universally-accepted solution.

This is a serious problem that needs to be solved before we can hope
to make rigorous testable predictions for an inflationary multiverse.
A vague hope is that somehow the dimensionality of the part of the
Hilbert space (or quantum state space if it is bigger than the Hilbert
space) where conscious observers are supported is finite, so that --
for all finite quantum states -- the expectation values of all finite
positive operators (including the awareness operators) would be
finite, thus giving finite measures for all conscious perceptions. 
But what would limit conscious observers to a finite dimensional part
of the presumably infinite-dimensional quantum state space eludes me.

\section{Testing multi-observation theories with\\typicality}

If we can find a theory that gives finite measures for sets of
observations (perhaps conscious perceptions) or which can be 
approximated as the expectation values of other positive operators,
how can we test it? If the theory predicts a unique observation (at
least unique under some condition, such as observing a clock reading
to have some value), then one can simply check whether one's
observation fits the prediction.  This would typically be the case in
a classical model of the universe with a single observer who reads a
clock that gives monotonically increasing readings (so that there is
only a single observation  for each clock reading).

Although a classical solipsist might believe this is true for his
universe, for most of us the evidence is compelling that there are
many observers and hence presumably many different observations even
at one value of some classical time.  Quantum theory further suggests
that there are many possible observations -- even for a single
observer at a single time.

There is a debate as to whether  the observations given by quantum
theory are actual or are merely unrealized possibilities. The
Copenhagen view seems to imply that -- for each value of the time and
for each observer -- there is only one observation that is actualized
(say by collapse of the wave-function), so that all the other
possibilities are unrealized.  This seems to come from a na\"{\i}vely
WYSIWYG\footnote{What You See Is What You Get (If you needed this
footnote, WYSIWYG is not WYSIWYG.)} view of the universe, so to me it
is much simpler to suppose that all possible observations predicted by
the quantum theory are actualized, with no ugly collapse of the
wave-function to give a single actualized observation for each
observer at each time.  We  are already used to the idea of many
different times (which are effectively just different branches of the
quantum state, at least in the Wheeler-DeWitt approach to quantum
gravity) and -- except for solipsists --  to the idea of many
different observers, so why should we not accept the simple prediction
from quantum theory of many observations at the same time by the same
observer?

In any case, whether in a classical universe or a quantum universe
without collapse of the wave-function, each time an observation
occurs, there are many observations even at the same time and so one
needs to be able to test this. To do this for a theory that gives
measures for all sets of observations, I would propose using the
concept of ``typicality''~\cite{SQM}, which is a suitable likelihood
that one may use to test or compare theories or to calculate their
posterior probabilities in a Bayesian analysis after assigning their
prior probabilities.

The basic idea is to choose a set of possible observations that each
give a single real parameter, such as the  Hubble constant or the
value of one of the constants of nature. Then we use the measure for
sets of observations to get the measure for all ranges of this single
real parameter.  For simplicity, we normalize the total measure in the
set of observations being considered to be unity.

Now we want to test one's observation against the theory by
calculating the typicality for that observation within the set.  For
simplicity, I shall call the observation being tested the `actual'
observation, even though the theory would say that all possible
observations with non-zero measure are realized as actual
observations. To do this, one calculates the total `left' and `right'
measures for all possible observations in the set under consideration,
i.e.  the total measure to the left or right of and including the
`actual' observation when they are ordered on the $x$-axis by the
value of the real parameter under consideration. These two measures
will add up to 1 plus the measure of the `actual' observation, which
is counted in both of the measures.

Next take the smaller of these two measures (the total measure on the
more extreme side if the `actual' observation is not in the middle of
the total measure) as the `extreme' measure of the `actual'
observation. We then use the normalized measure of the set of
observations to calculate the probability that a random observation
within the set would give an `extreme' measure as small as that of the
`actual' observation.  This probability is what I call the
`typicality' of the actual observation of the real parameter within
the chosen set of possible observations.  The typicality is thus the
probability that a random observation in the set is at least as
extreme as the actual observation.  It depends not only on the actual
observation but also on the theory predicting the measure for the sets
of observations. This is what is needed to calculate the conditional
probability of subsets of observations within the set under
consideration.

In the case in which the real parameter takes a continuum of values
and there is zero measure for an observation to have precisely any
particular value, the left plus right measures add up to unity. Then
the extreme measure (the smaller of the left and right measures) will
will take continuous values from 0 to $1/2$ with a uniform probability
distribution, so the typicality is twice the extreme measure.  In this
simple case, the typicality is a random variable with a uniform
probability distribution ranging from 0 (if the actual parameter is at
the extreme left or  right) to 1 (if the actual value is in the middle
of its measure-weighted range, with both left and right measures being
$1/2$).

If the real parameter takes on discrete values, then the situation is
more complicated.  For example, suppose that the real parameter is
$k$, with possible values $k=-1$ (with measure 0.2), $k=0$ (with
measure 0.35) and $k=+1$ (with measure 0.45).  Then $k=-1$ has a left
measure of 0.2 and a right measure of $0.2+0.35+0.45=1$ for an extreme
measure of 0.2; $k=0$ has a left measure of $0.2+0.35=0.55$ and a
right measure of $0.35+0.45=0.8$ for an extreme measure of 0.55; and
$k=+1$ has a left measure of $0.2+0.35+0.45=1$ and a right measure of
0.45 for an extreme measure of 0.45.  Thus the probability of an
extreme measure of 0.2 is 0.2 (the probability of $k=-1$); the
probability of an extreme measure of 0.45 is 0.45 (the probability of
$k=+1$); and the probability of an extreme measure of 0.55 is 0.35
(the probability of $k=0$).  The typicality of $k=-1$ is the
probability that the extreme measure will be at least as small as 0.2,
which is 0.2; the typicality of $k=0$ is the probability that the
extreme measure will be at least as small as 0.55, which is
$0.2+0.45+0.35=1$, and the typicality of $k=+1$ is the probability
that the extreme measure will be at least as small as 0.45, which is
$0.2+0.45=0.65$. 

Note that only for the most extreme parameter value or values (for
which the extreme measure is the smallest possible within the set) is
the typicality the same as the normalized measure of the observation
giving that value itself.  For less extreme parameter values, the
typicality is greater than the measure of the observations giving that
parameter value. On the other hand, the least extreme parameter value
or values (the middle one, for which the `extreme' measure is the
greatest possible within the set) has a typicality of unity.  Thus the
typicality always attains its upper limit of unity for some member of
the set, but the lowest value it attains is the measure of the most
extreme observation (which would be zero if the observed parameter
formed a continuum with zero measure for any particular value of the
parameter).

The typicality is thus a likelihood, given a theory for the measures
of sets of values of a real parameter, for a parameter chosen randomly
with the probability measure given by the theory, to be at least as
extreme as the `actual' observed parameter. The typicality has the
advantage over the probability measure for the actual observed
parameter of being a probability that has values up to unity for some
possible observation. This differs from the probability measure for
the parameter itself, which may have a very small upper limit (e.g. if
there is an enormous number of possible discrete values for the
parameter) or even a zero upper limit (e.g. if the parameter ranges
over a continuum and has a smooth probability density, with no delta
functions at any particular values of the parameters).

If one uses the probability measure itself as a likelihood, one cannot
directly do a Bayesian analysis with an observation of a continuous
parameter having a smooth probability density, since the resulting
likelihood will be zero for all possible observed values of the
parameter.  One might try to use the probability density instead of
the probability itself, but this depends on the coordinatization of
the parameter and so gives ambiguous results.  For example, one would
get a different likelihood for an observed value of the Hubble
constant $H$ by using its probability density than one would for
$H^2$.

Another approach that is often used for results that have a large
number of possible values is to bin them and then use the total
probability for the bin in which the actual observation lies as the
likelihood.  But again this depends on the bins and so gives ambiguous
results.   The ambiguity of both the probability density and the
binning are avoided if one uses the typicality as I have defined it
here.

Admittedly, if there are $N>1$ parameters being observed, then there
are ambiguities even with the typicalities. First, with more than one
parameter, one gets more than one typicality. Second, if there are $N$
independent parameters, one can construct $N$ independent combinations
of them in arbitrarily many different ways. Both of these problems are
related to the issue of how one chooses to test a theory, which has no
unique answer.

Once one has made a choice of what set of observations to include and
what parameter to determine the typicality for, how do we use the
typicality to test a theory?  It can be used -- like any other
likelihood -- in the following manner: Let $H_n$ be an  hypothesis
that gives measures to observations in the set, so that an actual
observation $O$ has typicality $T_n(O)$ according to this hypothesis. 
At the simplest level, one can say that, if $T_n(O)$ is low, then
$H_n$ is ruled out at the corresponding level.  For example, if
$T_n(O) < 0.01$, then one can say that $H_n$ is ruled out at the
$99\%$ confidence level.

A better approach would be to assign initial or prior probabilities
$P_i(H_n)$ to different hypotheses $H_n$, labeled by different values
of $n$.  Then the typicalities $T_n(O)$ for these different hypotheses
would be used as weights to adjust the  $P_i(H_n)$ to final or
posterior probabilities  $P_f(H_n)$ that are given by Bayes' formula:
\begin{equation}
P_f(H_n) = {T_n(O)P_i(H_n) \over \sum_m T_m(O)P_i(H_m)}\,.
\end{equation}
Apart from the ambiguity of choosing the set of possible observations
and the parameter to be observed and the physics problem of
calculating the typicalities $T_m(O)$ assigned by each theory $H_m$,
there is now the new ambiguity of assigning prior probabilities
$P_i(H_m)$ to the theories themselves.  This appears to be a purely
subjective matter, though -- in the spirit of Ockham's razor --
scientists would generally assign higher prior probabilities to
simpler theories.  Of course, there are arbitrarily many ways to do
that. However, if one just considered an infinite countable set of
theories that one could order in increasing order of complexity, from
the simplest $H_1$ to the next simplest $H_2$ and so on, then one
simple assignment of prior probabilities would be
\begin{equation}
P_i(H_m) = 2^{-m}\,.
\end{equation}

The idea of restricting attention to a countable set of theories seems
plausible, since humans could really consider only a finite set of
theories, but it could be inappropriate if the ultimate theory of the
universe contained an infinite amount of information, even if merely
in the form of a single real coupling constant or some other parameter
whose digits are not compressible (i.e. generated by a finite amount
of input information). Note that it is considered to be a merit of
string/M theory that there is  not even the possibility of having
infinite amounts of information in any dimensionless coupling
constants, at least in the dynamical equations of the theory, although
it is apparently not yet ruled out that the quantum state might have
an infinite amount of information. This might apply to the expectation
value of the dilaton, although most theorists would also prefer to
avoid this possibility.

\section{Testing the single-universe and multiverse\\hypotheses}

Tegmark \cite{Teg} has classified multiverse hypotheses into Levels 1,
2, 3 and 4. Level 1 is regions beyond our cosmic horizon, with the
same `constants of nature' as our own region.  Level 2 is other
post-inflation bubbles, perhaps with different `constants of nature'.
Level 3 are the Everett many worlds of quantum theory, with the same
features as Level 2.  Level 4 is other mathematical structures, with
different fundamental equations of physics as well as different
constants of nature.

Levels 1-3 can all come from a single universe if we define a universe
to be some quantum state in some quantum state space (e.g.\ some
C*-algebra state).  In this case, the quantum state-space may be
regarded as a set of quantum operators and their algebra, and the
quantum state as an assignment of an expectation value to each quantum
operator.  To get measures for observations in the form of conscious
perceptions, one must add to this bare quantum theory an assignment of
a particular positive operator for each set of conscious perceptions.
The resulting `awareness operators' then form a
positive-operator-valued set obeying the appropriate sum rules when
one forms unions of disjoint sets of conscious perceptions, so that
the resulting expectation values have the properties of a measure on
sets of conscious perceptions \cite{SQM,MS}.

Different hypotheses $H_m$ that each specify a single SQM universe 
would give different quantum state spaces, different operator
algebras, different quantum states, different sets of conscious
perceptions and/or different sets of awareness operators corresponding
to the sets of conscious perceptions.  (A quantum state is here
defined, in the C*-algebra sense, as the quantum expectation values
for  all possible quantum operators in the set.) By the SQM rule that 
the measure for each set of conscious perceptions is the expectation
value given by the quantum state for the corresponding awareness
operator, a definite SQM theory $H_n$ would give a definite measure
for each set of possible conscious perceptions.  This would be a
theory of a single SQM universe, though that universe could be a
multiverse in the senses of Levels 1-3.

Then by the procedure outlined above, from one's actual observation
$O$, a sufficient intelligence should be able to calculate for each
$H_m$ the typicality $T_m(O)$ of that observation. If one has a set of
such theories with prior probabilities $P_i(H_m)$, then one can use a
Bayesian analysis to calculate the posterior probability $P_f(H_n)$
for any specific theory $H_n$ and thereby test the theory at a
statistical or probabilistic level.

But what if there is more than one universe? Tegmark \cite{Teg,Teg2}
has raised the possibility of a multiverse containing different
mathematical structures, and it certainly  seems logically conceivable
that reality may consist of more than one universe  in the sense of 
Levels 1 to 3. Tegmark discusses a Level 4 multiverse which, as he
describes it, includes all mathematical structures.  This seems to me
logically inconsistent and  inconceivable. My argument against Level 4
is that different mathematical structures can be contradictory, and
contradictory ones cannot co-exist.  For example, one structure could
assert that spacetime exists somewhere and another  that it does not
exist at all.  However, these two structures  cannot both describe
reality.

Now one could say that different mathematical structures  describe
different existing universes, so that they each apply to separate
parts of reality and cannot be contradictory.  But this set of
existing universes, and the different mathematical structures with
their indexed statements about each of them, then forms a bigger
mathematical structure.  At the ultimate level, there can be only one
world  and, if mathematical structures are broad enough to include all
possible worlds or at least our own, there must be one unique
mathematical structure that describes ultimate reality. So I think it
is logical nonsense to talk of Level 4 in the sense of the
co-existence of all mathematical structures. However, one might want
to consider how to test levels of the multiverse between Levels 1-3
and 4.

One way to extend an SQM universe to a multiverse might be to allow
more than one quantum state on the same quantum state-space, while
keeping the other parts of the structure -- such as the awareness
operators -- the same.  Then if a weight is assigned to each of these
different quantum states, one can get the measure for each set of
conscious perceptions as the weighted sum of the measures for each
quantum state.  But this is equivalent to defining a new single
quantum state in a new single-universe theory that is the weighted sum
of these different quantum states in the original description. That
is, the  new single quantum state would be defined to give as the
expectation value of each operator the weighted sum of the expectation
values that the different quantum states would give.  (If the quantum
state can be described by a density matrix, then the new density
matrix would be the weighted sum of the old ones.)

Since the measure for a set of conscious perceptions in an SQM
universe is the expectation value given by the quantum state of the
awareness operator corresponding to the set of conscious perceptions,
one would get the same measure by using the new quantum state as by
taking the weighted sum of the measures in the old description in
which there are different quantum states.

Another way to get a broader multiverse would be to keep the same
quantum state-space, quantum operators, operator algebra and set of
possible conscious perceptions, but to include different sets of
awareness operators in different SQM universes.  But again, if one
weights the resulting measures for each universe to get a total
measure for this multiverse, this would be equivalent to forming a
single new set of awareness operators that are each the weighted sum
of the corresponding awareness operators in each of the different
universes.

Yet another way to extend the multiverse would be to include universes
with separate quantum state-spaces, each with its own quantum state
and awareness operators.  If each of these universes has a weight,
then one can again get the total measure for each set of conscious
perception by taking the weighted sum of the measure for that set in
each universe. This would be equivalent to defining a total quantum
state-space whose quantum operators were generated by the tensor sum
of the operators in each of the original sets of operators that
correspond to the original separate quantum state-spaces.  One could
take operators from different original sets as commuting to define the
quantum algebra of the new set.

The new single quantum state could then be defined by giving -- on any
sum of operators from the separate sets of operators -- the weighted
sum of expectation values that the old quantum states gave.  For
products of operators from different sets, one could just take the new
expectation value to be the product of the weighted old expectation
values for the separate operators in each set.  The new awareness
operators could be defined as the sum of the original awareness
operators.  Since this would involve only sums from the different sets
of operators and not products, the expectation values of the new
awareness operators would all be linear in the weights for the
original separate universes in the new single quantum state and hence
would give in that new single quantum state the same measure as the
weighted sum of the original measures.

Each of these three simple-minded ways to attempt to extend the
multiverse produces nothing new, at least for the measures of sets of
conscious perceptions.  Thus a single SQM universe is a fairly broad
concept, encompassing a wide variety of ways of generating measures
for conscious perceptions. In fact, one could argue that any
assignment of measures for conscious perceptions could come from a
single SQM universe, since one could just define awareness operators
for all sets of conscious perceptions and embed these into a larger
set of quantum operators with some algebra. One could then just choose
the quantum state to give the desired expectation values for all of
the awareness operators.

In principle, one could even choose the algebra of operators to be
entirely commuting, so that the resulting quantum theory would be
entirely classical, though still possibly giving the Everett many
worlds rather than just a single classical world.  Thus even a
universe that gives exactly the same measures for conscious
perceptions as ours, and hence the same typicalities for all
observations, could in principle be entirely classical in the sense of
being commutative.  We cannot prove that the universe is quantum just
from our observations.

However, surely such a classical description of our conscious
perceptions would involve a more complicated SQM universe than one in
which there are non-commuting operators (and presumably even
non-commuting awareness operators).  Thus it is on the ground of
simplicity and Ockham's razor that we assign higher probabilities to
non-commuting quantum theories that explain our observations, even
though the likelihoods for our observations can be precisely the same
in a classical theory. In a similar way it might turn out that,
although a multiple-SQM-universe theory could be reduced to a
single-SQM-universe theory in one of the ways outlined above, the
description could be simpler in terms of the former or even in terms
of universes that are not SQM.

If we do have a true multiverse of different universes, each of which
gives a measure for each set of conscious perceptions, then to get a
measure covering the whole of reality, we would need a measure  for
each of the individual multiverses.  For suppose each universe is
described by an hypothesis $H_n$ that assigns a measure $\mu_n(S)$ for
each set $S$ of conscious perceptions.  When we were considering
single universes, we considered different $H_m$ just as theoretical
alternative possibilities and discussed assigning subjective prior
probabilities $P_i(H_m)$ to them.  But when we are considering true
multiverses, we need an objective weight $w(H_n)$ for each universe,
since each universe with non-zero weight is being considered to
actually exist. Therefore the total measure for each set of conscious
perceptions from this extended multiverse would be $\mu(S) = \sum_n
w(H_n) \mu_n(S)$.

Extending the multiverse to multiple SQM universes (or to any ensemble
in which there is a prediction for the measure for all sets of
conscious perceptions from each universe) replaces our uncertainty
about which $H_n$ is correct with the uncertainty about which $w(H_n)$
is correct.  It would replace Tegmark's question \cite{Teg} ``Why
these equations?'' with ``Why this measure?'' We cannot evade some
form of this question by invoking ever higher levels of the
multiverse, even though  this may provide a simpler description of a
world.

In the sense that an SQM universe is a single universe, it may still
encompass Level 1-3 multiverses. At the true multiverse level, we need
not just a single theory $H_n$ for a single universe but also a
meta-theory $I$ for the measure or weight $w(H_n)$ of the single
universes within the set of actually existing multiverses. However,
since we do not yet know what the correct meta-theory is, just as we
do not yet know what the correct theory $H_n$ is for our single
universe, we may wish to consider various theoretically possible
meta-theories, $I_M$, labeled by some index $M$ in the same way that
$n$ labeled the single universe described by the theory $H_n$.  Then
meta-theory $I_M$ says that single  universes exist with measures
$w_{M,n} \equiv w_M(H_n)$ and so a set of conscious perceptions $S$
would have measure $\mu_M(S) = \sum_n w_{M,n} \mu_n(S)$.  From the
measure for conscious perceptions, one can follow the procedure
outlined in the previous section to get the typicality $T_M(O)$ of an
observation $O$ in meta-theory $I_M$.

For example, if the single universes described by $H_n$ are labeled
by the positive integers $n$ in order of increasing complexity,  and
if the meta-theories $I_M$ are labeled by the positive integers $M$,
one might imagine the following choice for the weights $w_{M,n}$ of
the meta-theory $I_M$ to give the universe $H_n$:
\begin{equation}
w_{2m-1,n} = {1\over m}\left({m\over m+1}\right)^n\,, \quad
w_{2m,n} = \delta_{mn}\,.
\label{mix}
\end{equation}
Then for odd $M$, one gets a geometric distribution of weights over
all single universes described by the theories $H_n$, with the mean of
$n$ being $m+1$. However, for even $M$, one gets a non-zero (unit)
weight only for the unique single universe described by the theory
$H_m$.  Thus the odd members of this countable sequence of
meta-theories do indeed give multiverse theories with various weights,
but the even members give single-universe theories.

Just as in a Bayesian analysis for single-universe theories we needed
subjective prior probabilities $P_i(H_m)$ for the possible
single-universe theories $H_m$, so now for a Bayesian analysis of
multiverse meta-theories $I_M$, we need subjective prior probabilities
$P_i(I_M)$. Again, although these subjective prior probabilities are
really arbitrary, we may wish to invoke Ockham's razor for the
meta-theories and assign the simpler ones the greater prior
probabilities. For example, if we can re-order the $I_M$ in increasing
order of complexity by another natural number $N(M)$, one might use
the simple subjective prior probability assignment
\begin{equation}
P_i(I_M) = 2^{-N(M)}\,.
\end{equation}
This would imply that the simplest meta-theory ($N=1$) is assigned
$50\%$ prior probability of being correct, the next simplest ($N=2$)
$25\%$ etc. 

For a more {\it ad hoc} choice, one could take the meta-theory weights
given by the hybrid model of eqn (\ref{mix}) for both single and
multiple universes and arbitrarily set
\begin{equation}
P_i(I_{2m-1}) = P_i(I_{2m}) = 2^{-m-1}\,.
\end{equation}
This gives a total prior probability of 1/2 for single-universe (even
$M$) theories and 1/2 for multiple-universe (odd $M$) theories. This
might be viewed as a compromise assignment if one is {\it a priori}
ambivalent about whether a single-universe or multiple-universe theory
should be used.

\section{Conclusions}

Even though multiverse theories usually involve unobservable elements,
they may give testable predictions for observable elements if they
include a well-defined measure for observations.  One can then analyze
them by Bayesian means, using the theory-dependent typicality of the
result of observations as a likelihood for the theory, though there is
still an inherent ambiguity in assigning prior probabilities to the
theories.

One can try to avoid specifying the equations or other properties of
an individual universe by assuming that there is an ensemble of
different universes, but this replaces the question of the equations
with the question of the measure for the different universes in the
ensemble.  There is no apparent way to avoid having non-trivial
content to a testable theory fully describing all of reality.

\section*{Acknowledgements}

I am very grateful for discussions with other participants at the 
Templeton conferences on the ``multiverse'' theme in Cambridge in 2001
and  Stanford in 2003.  My ideas have also been sharpened by e-mail
discussions with Robert Mann. This research has been supported in part
by the Natural Sciences and Engineering Research Council of Canada.

\end{document}